\documentclass[10pt,twocolumn]{article}
\usepackage{ol2}
\usepackage[draft]{hyperref}
\usepackage{amsmath}
\begin{document}

\twocolumn[ 

\title{Lensless ghost imaging with true thermal light}
\author{Xi-Hao Chen, Qian Liu, Kai-Hong Luo and Ling-An Wu}
\address{Laboratory of Optical Physics, Institute of Physics,
Chinese Academy of Sciences, Beijing 100190, China}

\begin{abstract}
We report the first (to our knowledge) experimental demonstration of
lensless ghost imaging with true thermal light. Although there is no
magnification, the method is suitable for all wavelengths and so may
find special applications in cases where it is not possible to use
lenses, such as with x-rays or $\gamma$-rays. We also show
numerically that some magnification may be realized away from the
focal plane, but the image will always be somewhat blurred.
\end{abstract}

] 

\noindent Since the first ``ghost" imaging experiment~\cite{Pittman}
based on quantum entangled photon pairs was performed, over the past
decade the phenomenon has attracted much attention in the field of
quantum optics. It is well-known now that ghost imaging, or
correlated two-photon imaging, can be performed not only with
entangled photon pairs but also with a classical thermal source. The
difference between these two approaches has been widely discussed by
the groups of Shih \cite{Valencia}, Boyd~\cite{Bennink2},
Lugiato~\cite{Gatti1}, Zhu~\cite{Zhu} and Wang~\cite{Cao}. More
recently, theoretical and experimental
studies~\cite{Cai,Scarcelli2,Scarcelli3,Basano,Ferri,Meyers,Meyers1,Han}
on lensless ghost imaging with thermal light have drawn new
attention; here ``lensless" means that no lens is used for imaging
the object. The possibility to perform lensless ghost imaging with
thermal light was first predicted by Wang and
collaborators~\cite{Cao}, who proposed in their paper that the
thermal source behaves as a phase-conjugate mirror which reflects an
object onto itself. The first experiment with a classical
pseudothermal source that successfully demonstrated lensless ghost
imaging was performed by Scarcelli et
al.~\cite{Scarcelli2,Scarcelli3}, which led to a
debate~\cite{Scarcelli2,Scarcelli3,Basano,Ferri,Gatti4,Scarcelli4}
on the question whether two-photon correlation phenomena must be
described quantum mechanically, regardless of whether the light
source is classical or quantum. Though there is still no consensus
on the subject so far, this does not affect potential applications
of ghost imaging with thermal light. In particular, the recent
lensless ghost imaging experiments in which reflected and scattered
light from the object were detected by second-order correlation
measurements show good promise for practical applications.
~\cite{Meyers,Meyers1} However, in all the above experiments, the
primary light source was pseudothermal radiation obtained by passing
a laser beam through a rotating ground glass plate.

\begin{figure}[htb]
\label{fig1} \centerline{\includegraphics[width=5cm]{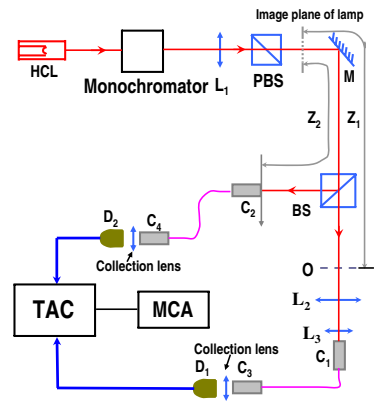}}
\caption{(Color online) Lensless ghost imaging experimental setup.
Object mask O is at a distance of $z_1=z_2=170$~cm from the image
plane of the lamp cathode; $\mathrm{D_1}$ is a bucket detector;
$\mathrm{C_2}$ is a fiber collimator scanned in the transverse
direction and connected to detector $\mathrm{D_2}$. See text for
explanation of other elements. }
\end{figure}

Different from these experiments and based on our previous
work~\cite{Zhangda,Zhai} with true thermal radiation, we report the
first demonstration of a lensless ghost imaging experiment using a
true thermal light source. We employed a commercial rubidium
hollow-cathode lamp (HCL) manufactured by the General Research
Institute for Nonferrous Metals (China); however, this time the
wavelength used was not the 780~nm of Rb as in our previous papers
but the 692.9~nm line of neon, a buffer gas in the lamp. The
coherence time $\mathrm{\tau_0}$ of the lamp powered by a direct
current of 30~mA was estimated from a Hanbury Brown-Twiss (HBT) type
measurement of the second-order correlation
function~\cite{Zhangda,hbtbook}, and found to be about 0.1~ns. This
is much shorter than that of experiments using pseudothermal
radiation from a laser beam randomly scattered by a ground glass
plate. An outline of the experimental setup is shown in Fig.~1. The
light from the lamp is passed through a monochromator to select out
the 692.9~nm spectral line, and is focused by the convex lens
$\mathrm{L_1}$ of 10~cm focal length to form a secondary light
source, an image of the cathode about 1.67~mm in diameter. A
polarizing beam splitter (PBS) transmits linearly polarized light.
After reflection by mirror M the beam is divided by a 50:50
nonpolarizing beamsplitter (BS). The object, a mask O consisting of
two pinholes of diameters 0.77 and 0.72~mm, 3.66~mm apart, is
inserted into the beam transmitted through BS. The mask was made
simply by pricking two holes in a piece of copper foil, with the
result that the hole diameters were not exactly equal. Two lenses
$\mathrm{L_2}$ and $\mathrm{L_3}$ act as a telescope, so that the
single-photon detector $\mathrm{D_1}$ (Perkin Elmer SPCM-AQR-12) can
capture all the light passing through the mask by means of fiber
collimators $\mathrm{C_1}$, $\mathrm{C_3}$ and a collection lens,
thus serving as a bucket detector. The reflected light from BS is
coupled into detector $\mathrm{D_2}$ by fiber collimators
$\mathrm{C_2}$, $\mathrm{C_4}$ and a collection lens. Note that the
lenses merely serve the purpose of collecting light; in the paths
from the effective plane of the lamp to the object and to fiber
collimator $\mathrm{C_2}$ ($z_1$ and $z_2$, respectively) there is
no lens. The receiving area of the collimators is about 1.8~mm in
diameter. The detector output signals are sent to a
time-to-amplitude converter (TAC), with detectors $\mathrm{D_1}$ and
$\mathrm{D_2}$ providing the ``start" and ``stop" signals,
respectively. The TAC output is connected to a multi-channel
analyzer (MCA), which displays a histogram of the coincidence counts
as a function of the difference in the times of arrival of the
photons at the two detectors.

 The transverse
normalized second-order correlation function is given
by~\cite{Valencia,Ferri}:
\begin{eqnarray} \label{g2 for ghost image}
g^{(2)}(x_2) \propto N + |T(x_2)|^2,
\end{eqnarray}
where $x_2$ is the transverse position of fiber collimator
$\mathrm{C_2}$, $T(x)$ the transmission function of the mask, and
$N$ the number of transparent features in the object, which equals 2
in our scheme because the mask has two pinholes. This equation
reflects the position-position correlation between the object and
image planes, as well as the fact that the visibility decreases
(background increases) when the number of points in the object
increases.

\begin{figure}[htb]
\label{fig2} \centerline{\includegraphics[width=6cm]{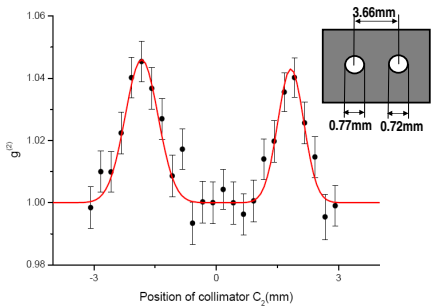}}
\caption{(Color online) Dependence of $g^{(2)}(x_2)$ on the position
of fiber collimator C$_2$, which gives the cross-sectional image of
the mask. The solid curve is a Gaussian fit. Inset: object mask with
two pinholes of diameters 0.77 and 0.72~mm, 3.66~mm apart.}
\end{figure}

In our experiment we choose the case in which the distance $z_1$
between the object and the effective plane of the lamp is equal to
the distance $z_2$ between the plane of the lamp and  the fiber
collimator $\mathrm{C_2}$, namely, $z_1=z_2=170$ cm. The collimator
$\mathrm{C_2}$ is scanned transversely across the reference beam in
steps of 0.25 mm, and the detector coincidence counts recorded. The
normalized second-order correlation function $g^{(2)}(x_2)$ is
calculated as previously~\cite{Zhangda,Zhai}, from which we plot the
cross-sectional image of the two-pinhole object, as shown in Fig. 2.
The two peaks are not symmetrical, clearly reflecting the slight
difference in size of the two pinholes, but it can be seen that
there is no magnification of the image. The visibility is found to
be $2.2\%$, which is lower than the value of $5\%$ that we obtained
in the HBT experiment. This is as expected, since different points
on an object will diminish the visibility of the image of other
points, so the more complicated the object, the worse will the
visibility be. Apart from the factor $N$ in expression (1), other
reasons for the lower visibility include the short coherence time
compared with the time jitter of the detection system, the limited
transverse coherence area of about 0.5 mm$^2$ in the object plane,
and the finite area of the fiber collimator $\mathrm{C_2}$. The
latter also lowers the resolution, which is, moreover, adversely
affected when the coherence area is too large.

We know that, in the classical optics approximation, a lens
generates an image of an object in the plane defined by the Gaussian
thin-lens equation. Basically, this equation defines a
point-to-point relationship between the object plane and image
plane. However, in practice it is sometimes said that an image can
also be obtained before and behind the focal plane, albeit at the
price of blurring the image. We would then expect that this could
also be said for second-order lens-focused ghost imaging with
thermal light, and therefore, also be true for the lensless scheme,
which is just a special case of having a lens of infinite focal
length. It has been reported in Ref. [13] that the longitudinal
coherence length of a thermal light source determines the region
where the ghost image exists. This implies that an image can be
obtained in any plane of this region, as in the above-mentioned case
of classical first-order imaging. However, in the out-of-focus
condition $z_1\neq z_2$ there will never be a ``perfect" image. This
has been seen from the experimental results in Ref. [14], where a
perfectly sharp image is only obtained in the focused condition
$z_1=z_2$ and the imaging quality becomes worse and worse with the
increase of $\Delta z$ ($=z_2-z_1$), as in the classical case. Thus
strictly speaking there is no magnification in lensless ghost
imaging; an image in the focal plane has the same size as the
object, while all images in the out-of-focus planes are blurred
compared with that at the focus.

For a better understanding of lensless ghost imaging in the
defocused case we perform a numerical simulation of an experiment.
The experimental scheme is almost the same as in some of the papers
mentioned above. With $x_1$ and $x_2$ representing the transverse
coordinates in the planes of the bucket detector $\mathrm{D_1}$ and
the point detector $\mathrm{D_2}$, the transverse normalized
second-order intensity fluctuation correlation function is
\begin{eqnarray} \label{g2}
\Delta g^{(2)}(x_2)= \int\frac { \langle\Delta I_1(x_1) \Delta
I_2(x_2) \rangle }{\langle I_1(x_1)\rangle \langle I_2(x_2)
\rangle}dx_1 ,
\end{eqnarray}
where $\Delta I_i(x_i)$ and $I_i(x_i)$ are the intensity fluctuation
and intensity at the detector positions $x_i ~(i=1, 2)$,
respectively. Thus we see that $\Delta g^{(2)}(x_2)$ depends on the
second-order intensity fluctuation correlation $\langle\Delta
I_1(x_1) \Delta I_2(x_2) \rangle$, which can be expressed as
\begin{eqnarray}\label{}
\langle\Delta I_1(x_1) \Delta I_2(x_2) \rangle \nonumber
\propto\left|{\int{h_2^{*}(x^{'},x_2)h_1(x^{'},x_1)dx^{'}}}\right|^2\\
 \propto  \left|{\int{\mathrm{exp}[\frac{{\pi ^2 (x^{'}- x_1 )^2 (x^{'} - x_2 )^2 }}{{\lambda ^2
z_1 z_2 }}]T(x_1 )dx^{'}} } \right|^2 ,
\end{eqnarray}
in which
\begin{eqnarray}\label{}
h_1(x,x_1) &\propto& \mathrm{exp}{\rm{[}}\frac{{i{\rm{\pi (}}x -
x_{\rm{1}} {\rm{)}}^{\rm{2}} }}{{\lambda z_{\rm{1}} }}{\rm{]
}}T{\rm{(}}x_{\rm{1}} {\rm{)}},
\end{eqnarray}
\begin{eqnarray}\label{}
 h_2(x,x_2) &\propto&
\mathrm{exp}[\frac{{i\pi (x - x_2 )^2 }}{{\lambda z_2 }}]{\rm{ }}  ,
\end{eqnarray}
and $h_i(x,x_i)$ is the impulse response function for light
propagating from a point $x$ on the source to a point $x_i$ in the
detector plane, and $z_1$, $z_2$ are the respective distances from
the source to the object and to the detector $\mathrm{D_2}$.

In our simulation a light source of wavelength 693~nm and radius
6~mm and $z_1$=300~mm is chosen, and the object is a double-slit
with slits of width 100~$\mu$m separated by 200~$\mu$m. The
normalized second-order intensity fluctuation correlation function
$\Delta g^{(2)}$ is plotted against $x_2$ and $z_2$ in Fig.~3. It
can be clearly seen that both the visibility and the resolution of
the image become worse and worse as $z_2$ deviates from the position
$z_2=z_1$, and a sharp image can only be obtained in the focused
condition. Though we can obtain fairly clear images a short distance
away, they are still blurred compared with that in the plane of
focus. It is true that the images in the planes of $z_2<z_1$ are
broader than at the focus, but it would be rather factitious to say
that this is actually magnification. It is also contradictory to say
that the image magnification is $z_2/z_1$, as mentioned in [10] and
[13], since in this case we have $z_2<z_1$.

\begin{figure}[htb]
\label{fig3}
\centerline{\includegraphics[width=7.0cm]{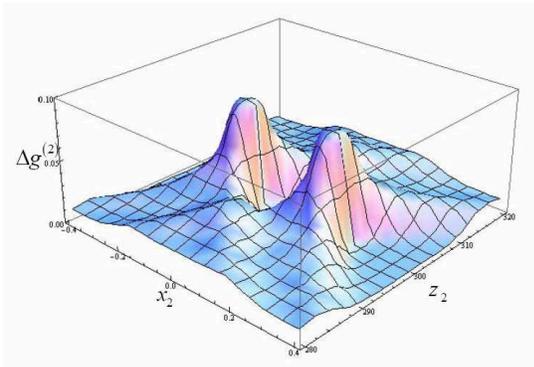}}
\caption{(Color online) Numerical simulation of $\Delta g^{(2)}$ for
lensless ghost imaging as a function of $x_2$ and $z_2$ in mm. The
object is a double-slit of slit width $100~\mu$m and separation $
200~\mu m$, at a distance of $z_1$=300~mm from a 693~nm thermal
light source of radius 6~mm.}
\end{figure}

In conclusion, we have experimentally realized lensless ghost
imaging with true thermal light. Although the original visibility is
very low the reconstructed image may be clearly recognizable as the
background can be easily removed by standard image processing
techniques. Since the imaging setup works for any wavelength and no
lenses are required, such a method seems quite promising for imaging
applications at wavelengths such as x-rays or $\gamma$-rays where no
effective lens is available. Thermal light sources are easier to
obtain, and so it is conceivable that they could find certain
special applications~\cite{ShenshengHan} where entangled or
pseudo-thermal sources are not so convenient to use. We have also
demonstrated theoretically that in thermal lensless ghost imaging a
really sharp image can only be obtained at the focus, in which case
there is no magnification. The image in all other planes even when
within the longitudinal coherence length is always blurred compared
to that in the focal plane.

This work was supported by the National Natural Science Foundation
of China (Grants 60578029 and 10674174), and the National Program
for Basic Research in China (Grant 2006CB921100). L. A. Wu's e-mail
address is wula@aphy.iphy.ac.cn.

\end{document}